\documentclass[prl,superscriptaddress,twocolumn,longbibliography]{revtex4-1}

\usepackage{amsmath}
\usepackage{amssymb}
\usepackage{amsxtra,color}
\usepackage{latexsym}
\usepackage{graphicx}
\usepackage{tikz}
\usepackage{pgfplots}
\usepgfplotslibrary{groupplots}
\pgfplotsset{compat=1.3}
\usepackage{subcaption}
\usepackage{lipsum}
\usepackage{color}
\usepackage{soul}

\usepackage[utf8]{inputenc}
\usepackage{MnSymbol}	

\usepackage{hyperref}

\definecolor{MyDarkGreen}{rgb}{0,0.6,0}
\definecolor{MyDarkBlue}{rgb}{0,0,0.8}
\definecolor{MyDarkRed}{rgb}{0.6,0,0.3}

\hypersetup{breaklinks=true, colorlinks=true,plainpages=true, linktocpage=true, linkcolor=MyDarkBlue, citecolor=MyDarkGreen, urlcolor=MyDarkRed, pdfborder={0 0 0},%
pdfauthor={},%
pdfsubject={Research article},
pdftitle={}%
}

\newcommand{\ba}{\begin{align}}
\newcommand{\ea}{\end{align}}
\newcommand{\be}{\begin{equation}}
\newcommand{\ee}{\end{equation}}

\newcommand{\braOket}[3]{\langle \: #1 \: | \: #2 \:| \: #3 \: \rangle}

\makeatletter
\newcommand*{\rom}[1]{\expandafter\@slowromancap\romannumeral #1@}
\makeatother
\begin{document}

 \title{Fano's propensity rule in angle-resolved attosecond pump--probe photoionization}

\author{David \surname{Busto}}
\affiliation{Department of Physics, Lund University, Box 118, SE-221 00 Lund, Sweden}

\author{Jimmy \surname{Vinbladh}}
\affiliation{Department of Physics, Stockholm University, AlbaNova University Center, SE-106 91 Stockholm, Sweden}

\author{Shiyang \surname{Zhong}}
\affiliation{Department of Physics, Lund University, Box 118, SE-221 00 Lund, Sweden}

\author{Marcus \surname{Isinger}}
\affiliation{Department of Physics, Lund University, Box 118, SE-221 00 Lund, Sweden}

\author{Saikat \surname{Nandi}}
\affiliation{Department of Physics, Lund University, Box 118, SE-221 00 Lund, 
Sweden}

\author{Sylvain \surname{Maclot}}
\affiliation{Department of Physics, Lund University, Box 118, SE-221 00 Lund, 
Sweden}
\affiliation{Biomedical and X-Ray Physics, Department of Applied Physics, AlbaNova University Center, KTH Royal Institute of Technology, SE-10691 Stockholm, Sweden}

\author{Per \surname{Johnsson}}
\affiliation{Department of Physics, Lund University, Box 118, SE-221 00 Lund, 
Sweden}

\author{Mathieu \surname{Gisselbrecht}}
\affiliation{Department of Physics, Lund University, Box 118, SE-221 00 Lund, Sweden}

\author{Anne \surname{L'Huillier}}
\affiliation{Department of Physics, Lund University, Box 118, SE-221 00 Lund, Sweden}

\author{Eva \surname{Lindroth}}
\affiliation{Department of Physics, Stockholm University, AlbaNova University Center, SE-106 91 Stockholm, Sweden}

\author{Jan~Marcus \surname{Dahlstr\"om}}
\email{marcus.dahlstrom@matfys.lth.se}
\affiliation{Department of Physics, Lund University, Box 118, SE-221 00 Lund, Sweden}


\begin{abstract}

In a seminal article, Fano predicts that absorption of light occurs preferably with increase of angular momentum. Here we generalize Fano’s propensity rule to laser-assisted photoionization, consisting of absorption of an extreme-ultraviolet photon followed by absorption \textit{or} emission of an infrared photon. The predicted asymmetry between absorption and emission leads to incomplete quantum interference in attosecond photoelectron interferometry. It explains both the angular-dependence of the photoionization time delays and the delay-dependence of the photoelectron angular distributions. Our theory is verified by experimental results in Ar in the 20-40 eV range.

\end{abstract}

\maketitle 

In quantum mechanics, the possible transitions between different states are dictated by {\it selection rules} which are based on symmetry arguments. For example, the famous parity and angular momentum selection rules are at the core of our understanding of light-matter interactions. In contrast to these stringent selection rules, the concept of {\it propensity rules}, which was introduced by Berry \cite{BerryJChemPhys1966}, is based on the fact that all allowed transitions are not equally probable. Selection and propensity rules play a fundamental role in physics and chemistry to understand reaction outcomes and probabilities. One of the most fundamental reactions is the photoionization of an atom following the absorption of a high energy photon, $\mathrm{A} +\gamma \rightarrow \mathrm{A}^+ +e^-$, where an electron is promoted to a manifold of degenerate continuum states. The well-known electric dipole selection rules greatly simplify the problem, restricting the possible transitions to those where the electron angular momentum changes by one unit, $\Delta \ell=\pm 1$. Fano's propensity rule states that out of the two possible transitions, the one increasing the electron angular momentum is favored,  due to increase of the centrifugal potential with angular momentum \cite{FanoPRA1985}.

This Letter deals with laser-assisted photoionization, 
$\mathrm{A} +\gamma_\textsc{XUV} \pm \gamma_\textsc{IR} \rightarrow \mathrm{A}^++e^-$, where absorption of an extreme ultraviolet (XUV) photon brings an electron to the continuum, followed by absorption or stimulated emission of infrared (IR) laser radiation between continuum states.
Laser-assisted photoionization is a cornerstone of attosecond science, used in the temporal characterization of XUV radiation, such as high-order harmonics  \cite{GloverPRL1996,SchinsJPB1996,MauritssonPRA2004} and the measurement of attosecond pulses using the reconstruction of attosecond beating by interference of two-photon transitions (RABBIT) scheme \cite{VeniardPRA1996,PaulScience2001}.  This technique is used in many applications, especially the investigation of photoionization dynamics in atoms \cite{SchultzeScience2010,KlunderPRL2011,IsingerScience2017,KoturNatCommun2016,GrusonScience2016}, molecules \cite{HaesslerPRA2009,HuppertPRL2016,BeaulieuScience2017,CattaneoScience2018} and solids \cite{CavalieriNature2007,LocherOptica2015,KasmiOptica2017}. 
 A natural question is whether Fano's propensity rule, originally stated between bound and continuum states, can be extended to laser-assisted photoionization and how a possible asymmetry between absorption and emission affects attosecond photoemission measurements. 

 \begin{figure*}[tb!]
  \includegraphics[width=\textwidth]{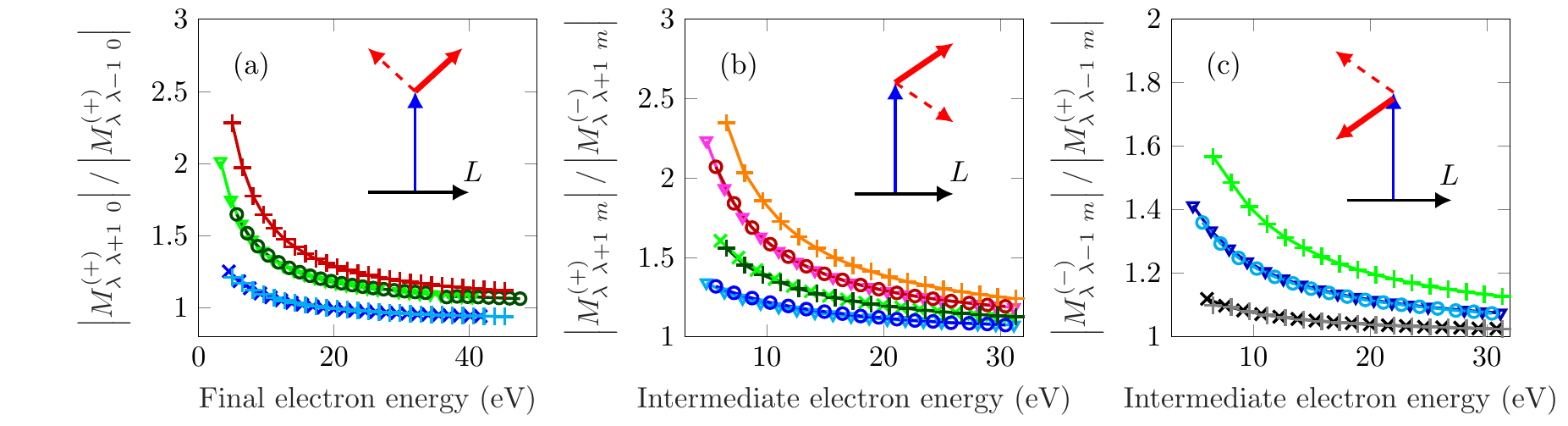}
  \caption{\textbf{Propensity rules in laser-assisted photoionization} in He $1s$ ($\times$), Ne $2p$ ($\triangledown$), Ar $3p$ (\small{$\bigcirc$}) and Kr $3d$ ($+$). The color of the curves indicates the angular momentum of the intermediate state in (a) and final state in (b), (c) [shades of grey $s$, shades of blue $p$, shades of green $d$, shades of red $f$, orange $g$]. (a) Probability ratio between increasing ($\lambda\rightarrow\lambda+1$) and decreasing angular momentum ($\lambda\rightarrow\lambda-1$) in the case of absorption of a photon from the intermediate state. (b) Probability ratio between absorbing and emitting a photon in the case of increasing angular momentum. (c) Probability ratio between emitting and absorbing a photon in the case of decreasing angular momentum. The insets present an energy and angular momentum diagram illustrating the propensity rule in each case.}
\label{Fig:1}
\end{figure*}
 
Recently, attosecond experimental observations have been extended to include angle resolution, with the aim to provide information on multiple competing channels \cite{VilleneuveScience2017}. These studies show that the measured photoionization time delays strongly depend on the electron emission angle \cite{HeuserPRA2016,CirelliNatCommun2018}. Conversely, the photoelectron angular distributions (PAD) vary as a function of the delay between the attosecond pulse train and the probe field \cite{AseyevPRL2003,CirelliNatCommun2018}. Modifications of PADs with pump-probe delay have been theoretically predicted \cite{GramajoPRA2017,GuyetandJPB2005} and the angular dependence of the photoionization time delays in RABBIT measurements has been reproduced in several numerical calculations \cite{HeuserPRA2016,IvanovPRA2017,BrayPRA2018,DahlstromJPB2014,WatzelJPB2014,HockettJPB2017}. Nonetheless, the underlying physical origin of these effects is still not understood.

Here, we demonstrate that Fano's propensity rule can be generalized to laser-assisted photoionization in atoms. Subsequently, we show that the asymmetry between absorption and emission leads to \textit{incomplete} quantum interference in the RABBIT scheme which is at the origin of the delay-dependence of the PADs and the angle-dependence of the photoionization time delays. Finally, we verify our theoretical predictions by comparing with experimental results obtained in argon.

Our theoretical approach is based on angular-channel-resolved many-body perturbation theory. The two-photon transition matrix elements corresponding to absorption of the $(s\mp 1)^ \mathrm{th}$ harmonic followed by absorption or emission of a fundamental IR photon with angular frequency $\omega$  are calculated according to (we use atomic units: $\hbar =m=e=4\pi\varepsilon_0=1$ unless stated otherwise)
\begin{align}
M_{\lambda Lm}^{\mathrm{(\pm)}}= \lim_{\epsilon\rightarrow0^+}\sumint_p \ 
\frac{\braOket{q}{z}{p}\left\langle p\right|z+\delta z\left|a\right\rangle }{\varepsilon_{a}-\varepsilon_{p}+\omega(s\mp1)+i\epsilon},
\label{Eq:MLm}
\end{align}
where $a$ is the initial state, with energy $\varepsilon_a$, $q$ the final electron continuum state, with angular momentum $L$, and $p$ the intermediate states, with energy $\varepsilon_p$ and angular momentum $\lambda$. The $(\pm)$ sign refers to absorption or emission of the IR photon. The common linear polarization of XUV and IR fields, along the $z$ axis, ensures conservation of the magnetic quantum number, $m$. The amplitudes of the XUV and IR fields are set to one for simplicity. 
Our calculations are based on a one-electron Hamiltonian, with a Dirac-Fock potential plus a correction that ensures the correct long-range potential for ionized photoelectrons  \cite{DahlstromJPB2014}. Electron correlation effects are included by self-consistent changes in the interaction, $\delta z$, known as the Random Phase Approximation with Exchange (RPAE), for absorption of the XUV photon \cite{AmusiaAtomicPhotoeffect1990}. 

Figure~\ref{Fig:1}(a) presents ratios between two-photon matrix elements corresponding to increasing or decreasing angular momentum in the IR absorption process, from the same intermediate state in the continuum. The ratios, calculated for various noble gas atoms ionized from different shells [outer shell of helium ($1s$), neon ($2p$) and argon ($3p$), or inner-shell of krypton ($3d$)], are generally larger than one, which means that IR absorption is more probable towards higher angular momentum. These ratios are to a large extent atom-independent, only showing a dependence on the final electron energy and the intermediate angular momentum. Figure \ref{Fig:1} also compares absorption and emission from the same intermediate continuum state. The absorption/emission ratio is larger than one when angular momentum is increased, $\lambda\rightarrow\lambda\!+\!1$ [Fig.~\ref{Fig:1}(b)], while the emission/absorption ratio is larger than one when angular momentum is decreased, $\lambda\rightarrow\lambda-1$ [Fig.~\ref{Fig:1}(c)], thus reflecting the time-reversal symmetry of the laser-driven continuum transition.

Similar to the case of one-photon transitions between bound and continuum states, discussed by Fano \cite{FanoPRA1985}, the physical origin of the propensity rule can be understood by evaluating the radial part of the transition matrix elements. The strength of the transition between the intermediate and final states is largest when the difference between the local momenta $k(r)=\sqrt{2[E-V(r)]}$ is smallest. (Here $E$ is the electron kinetic energy and $V(r)$ the radial potential). Hence, in the case of absorption (emission) of a photon, increasing (decreasing) angular momentum is favored because the increased (decreased) kinetic energy of the electron is compensated by a larger (smaller) centrifugal potential. In addition, the asymmetry between absorption and emission increases with angular momentum, since the centrifugal potential varies as $L(L+1)$. 

It should be pointed out that there is also a competing effect, due to the angular momentum part of the transition matrix elements, which always favours lowering the angular momentum. Indeed, in Fig.~\ref{Fig:1}(a), at energies larger than 20 eV, Fano's propensity rule is violated in the case of absorption of a photon from an intermediate $p$ continuum. In this case, the centrifugal potential is much smaller than the electron kinetic energy, leading to a small radial effect. Here the angular effect takes over, leading to a ratio slightly lower than one. Conversely, the ratios over different final angular momenta for emission of a photon in the continuum show larger asymmetry, because both radial and angular effects favour the decrease of angular momentum (not shown). 
In Fig.~\ref{Fig:1}(b) and (c) the ratios are always larger than one because final states with the same angular momentum are compared.

The RABBIT technique relies on the interference of two-photon transitions using an XUV attosecond pulse train, consisting in a comb of odd-order harmonics and a weak phase-locked IR field \cite{PaulScience2001}. The strength of the created sideband (SB), $P_\mathrm{SB}$, photoelectron peaks oscillates with the delay, $\tau$, between the XUV and IR fields according to
\begin{equation}
    P_{\mathrm{SB}}(\tau)=a+b\cos{\left(2\omega\tau+\delta\varphi\right)}.
    \label{Eq:RABBIT_equation}
\end{equation}
The measured photoionization time delay is defined as $\tau_A=-\delta \varphi/2\omega$ \cite{KlunderPRL2011}. In the case of angle-resolved measurements, the coefficients $a$, $b$ and the phase offset $\delta\varphi$ depend on the emission angle $\theta$ \cite{HeuserPRA2016,CirelliNatCommun2018}. 

Any angular distribution with azimuthal symmetry can be described as 
 \begin{equation}
 S(\theta)=\frac{\sigma_0}{4\pi}\left[1+\sum_{n=1}^{\infty} \beta_n P_n(\cos \theta)\right],
 \label{Eq:beta_parameters1}
\end{equation}  
where $P_n$ are Legendre polynomials, $\beta_n$ are the so-called asymmetry parameters, and $\sigma_0$ is the total cross-section. For two-photon transitions, the sum can be stopped at $n=4$ and $\beta_1=\beta_3=0$ because the final states have the same parity, giving rise to an up-down symmetric PAD described by $\beta_2$ and $\beta_4$ \cite{ArnousPRA1973,ReidAnnuRevPhysChem2003}. 
In the case of RABBIT measurements, the sideband PAD, $P_{\mathrm{SB}}(\theta,\tau)$, can be described by Eq.~(\ref{Eq:beta_parameters1}), with delay-dependent asymmetry parameters, $\beta_2(\tau)$, $\beta_4(\tau)$. Using a partial wave expansion, the angle-resolved SB signal can be formally expressed as  
\begin{equation}
\!\!P_{\mathrm{SB}}(\theta,\tau)\!=\!\!\!\int\!\!\!\! \mathrm{d}\phi\sum_m\big|\sum_{L} Y_{Lm}(\theta,\phi)e^{-i\frac{L\pi}{2}+i\eta_L}
A_{Lm}(\tau)\big|^2,
\label{Eq:ISB}
\end{equation}
where $\phi$ is the azimuthal angle, $\eta_L$ the phase of the scattering state with angular momentum $L$ and 
$Y_{Lm}(\theta,\phi)$ the spherical harmonics. The amplitudes $A_{Lm}(\tau)$ are given by
\begin{equation}
 \!\!A_{Lm}(\tau)=\!\!\sum_{\lambda} \Big[M_{\lambda Lm}^{(+)}e^{i\omega\tau} + M_{\lambda Lm}^{(-)}e^{-i\omega\tau}\Big].   
\label{Eq:alm}
\end{equation}
Using an asymptotic approximation described in \cite{dahlstromCP2013}, the phase of the transition matrix elements can be simplified such that
\begin{equation}
\arg(M_{\lambda Lm}^{(\pm)}e^{-i\frac{L\pi}{2}+i\eta_L})\approx\eta_\lambda^{(\pm)} +\phi_{cc}^{(\pm)} - (\lambda+1)\tfrac{\pi}{2},
\label{Eq:phaseM}
\end{equation}
where $\phi_{cc}^{(\pm)}$ represents the phase of the continuum-continuum transition and $\eta_\lambda^{(\pm)}$ is the scattering phase of the intermediate state. 

We now consider RABBIT in He, where only one intermediate angular momentum ($\lambda=1$) is reached and where $m=0$. 
  In this case, Eq.~(\ref{Eq:ISB}) simplifies to
\begin{equation}
\!\!P_{\mathrm{SB}}(\theta,\tau)\!=\!2\pi|A_{20}(\tau)e^{i\eta_2}Y_{20}(\theta)-A_{00}(\tau)e^{i\eta_0}Y_{00}| ^2. 
\label{Eq:psbhe}
\end{equation}
In angle-integrated measurements, the orthogonality of the spherical harmonics leads to an incoherent sum of the two final momenta contributions ($L=0,2$) in Eq.~(\ref{Eq:psbhe}). Since there is only one intermediate angular momentum, these terms will oscillate in phase provided that the continuum--continuum phases, $\phi_{cc}^{(\pm)}$, do not depend on $L$. Hence, a unique photoionization time delay can be accurately extracted \cite{KlunderPRL2011,dahlstromCP2013}.

We now consider angle-resolved RABBIT. At high kinetic energy, referred to as the soft-photon regime \cite{MaquetJModOpt2007}, absorption and emission amplitudes are symmetrical, i.e. $|M^{(+)}_{1L0}| \approx |M^{(-)}_{1L0}|$. The $\tau$ and $\theta$ dependencies in Eq.~(\ref{Eq:psbhe}) can then be separated into two independent factors. Consequently, the delay dependence of the PAD and the angle dependence of the photoionization time delay disappear. 

\begin{figure}[tb!]
 \resizebox {\columnwidth} {!} {
\includegraphics[]{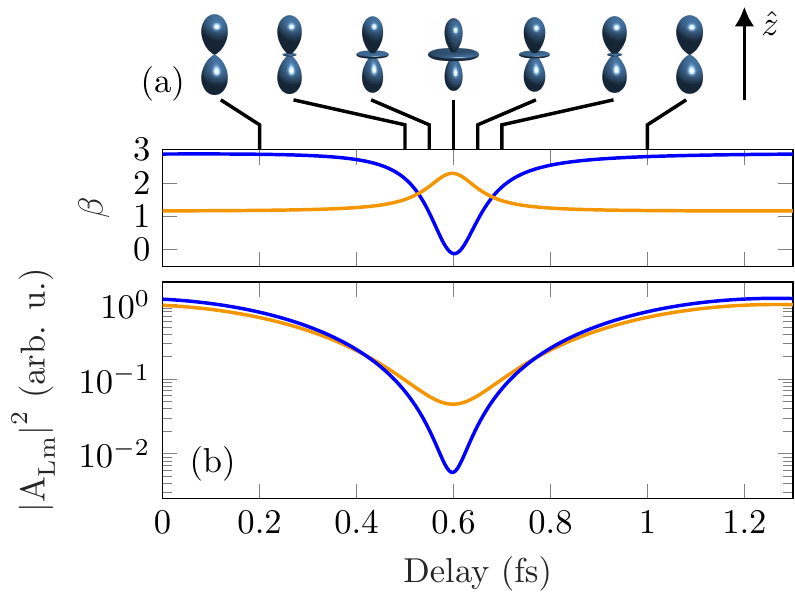}
 }
\caption{\textbf{Delay dependence of the PAD.} (a) Asymmetry parameters $\beta_2$ (blue) and $\beta_4$ (yellow) as a function of the delay for SB 20 in He. The pictures at the top show the evolution of the PADs together with polarization direction $\hat{z}$. (b) Temporal dependence of $|A_{20}|^2$ (yellow) and $|A_{00}|^2$ (blue) normalized to the maximum of $|A_{00}|^2$. }
\label{Fig:2}
\end{figure}
At lower kinetic energy ($\lesssim 30$ eV), previous approximations are not valid, and a rigorous treatment requires careful account of interfering terms in Eq.~(\ref{Eq:psbhe}). Figure \ref{Fig:2}(a) shows the evolution of the asymmetry parameters of SB 20 in He for delays spanning one SB oscillation period as well as the corresponding PADs at specific delays. The PAD varies strongly as a function of the delay, with a periodic emission of electrons perpendicular to the laser polarization. This can be understood by comparing the relative weight of the $s$ and $d$ contributions, $|A_{L0}(\tau)|^2$, as a function of delay [Fig.~\ref{Fig:2}(b)]. Around the SB maximum, the two contributions are comparable, while at the SB minimum, the $s$ state contribution drops significantly and the $d$ state becomes dominant, resulting in a strong modification of the PAD. 
The asymmetry between absorption and emission leads to incomplete quantum interference and reduction of the contrast of the oscillations ($|A_{L0}|^2$ remains different from zero). This effect is stronger for the $d$ contribution for the $s$ one, as predicted by Fano's propensity rule.

\begin{figure}[tb!]
\includegraphics[]{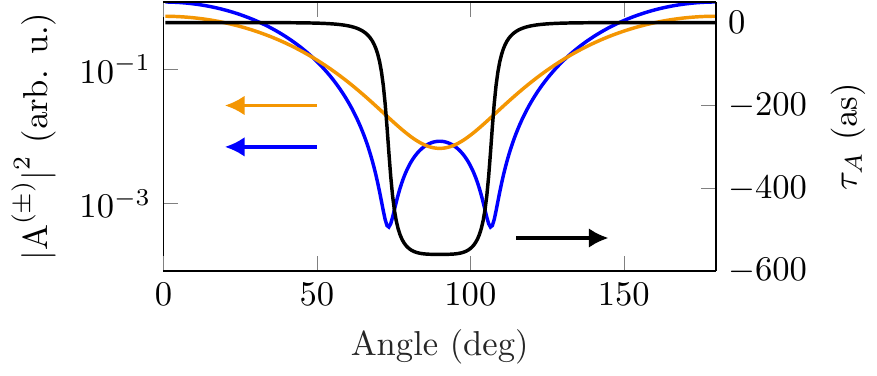
}
\caption{\textbf{Angular dependence of the photoionization time delay}: Contributions of the absorption ($A^{(+)}$, blue) and emission ($A^{(-)}$, yellow) paths and atomic delay ($\tau_A$, black) in SB 14 in He.}  
\label{Fig:PhaseHe}
\end{figure}
To understand the origin of the angular dependence of the photoionization time delays, we introduce the angle-resolved contribution of the absorption (emission) path
\begin{equation}
\!\!\!\!A^{(\pm)}(\theta)\!\approx\!e^{i[\pm\omega\tau+\eta_\lambda^{(\pm)}+\phi_{cc}^{(\pm)}]}\left[\left| M_{120}^{(\pm)}\right|Y_{20}(\theta)\!-\!\left|M_{100}^{(\pm)}\right|Y_{00}\right].
    \label{Eq:AThetaHe}
\end{equation}    
Depending on the relative weight of the $L$-dependent amplitudes, the $s$ and $d$ contributions may interfere destructively at specific angles, leading to a $\pi$ phase jump of the argument of $A^{(\pm)}(\theta)$. As shown in Fig.~\ref{Fig:PhaseHe}, the absorption path to SB 14 presents strong destructive interference at $75^\circ$ and $105^\circ$ indicating that $A^{(+)}(\theta)$ undergoes a $\pi$ phase jump [see supplemental material (SM)]. In contrast, the emission path does not exhibit any destructive interference and varies smoothly as a function of angle. This behaviour also results from the asymmetry between absorption and emission explained by the generalized Fano's propensity rule. 

In Fig. \ref{Fig:PhaseHe}, the calculated atomic delay does not depend much on the emission angle, except at the positions close to the destructive interference in the absorption path. In exact numerical calculations, a slight $L$-dependence of $\phi_{cc}^{(\pm)}$ occurs at low energy \cite{dahlstromCP2013}. This smooths the variation of the time delay around 75$^\circ$ and 105$^\circ$, leading to a phase jump  smaller than $\pi$ as shown in Fig. \ref{Fig:PhaseHe}. (A $\pi$ phase jump corresponds to a delay of 667 as.) As the kinetic energy increases, the angles at which the destructive interference occur move towards 90$^\circ$ since the asymmetry between absorption and emission becomes weaker and the phase jump closer to $\pi$ (see SM). 

\begin{figure}[tb!]
 \resizebox {\columnwidth} {!} {
\includegraphics[]{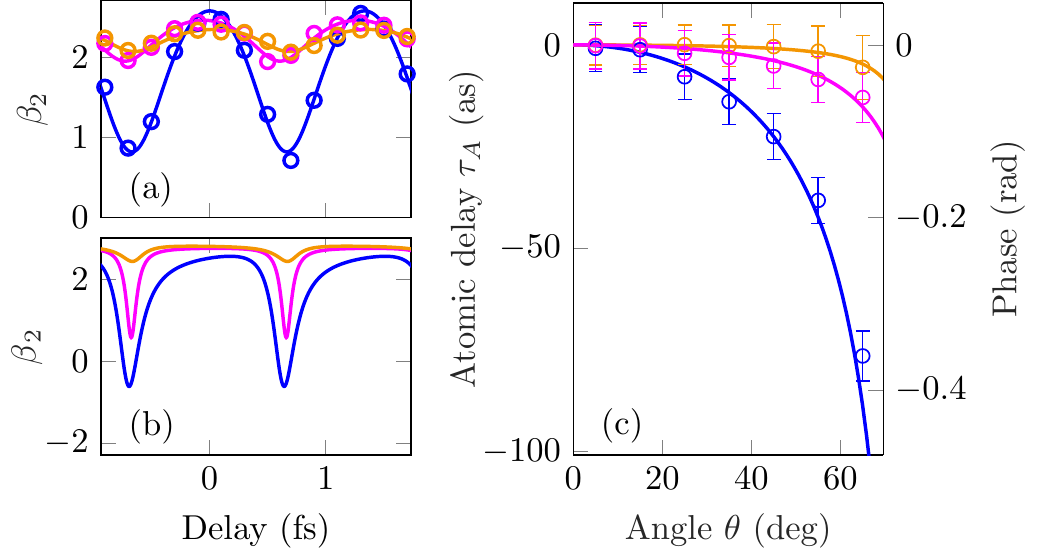}
 }
\caption{\textbf{Experimental and theoretical results in Ar} : (a) Experimental and (b) theoretical variation of $\beta_2$ as a function of delay for SBs 14 (blue),
20 (magenta) and 22 (yellow). Solid lines in (a) are fits to the data. (c) Experimental (circles) and theoretical (solid curves) angle dependence of the atomic delay for SBs 14, 20 and 22. The error bars correspond to the standard deviation returned from the fitting algorithm.}
\label{Fig:Ar}
\end{figure}

We now study the case of Ar, for which we present both experimental measurements and theoretical calculations of the asymmetry parameter $\beta_2$ as a function of the delay as well as angle-resolved photoionization time delay measurements. This case is more complex than He since there are three incoherent channels corresponding to $m=0,\pm1$ and since for $m=0$, there are two possible intermediate angular momenta ($\lambda=0,2$). In the experiments, we focus a XUV attosecond pulse train with  photon energy in the 20-40~eV range, together with a fraction of the fundamental IR field ($\hbar\omega=1.58$\,eV) with a variable delay, into a gas jet of argon atoms. The photoelectrons are detected using a Velocity Map Imaging Spectrometer (see \cite{CirelliNatCommun2018,RadingAppliedSciences2018} for more details on the experimental setup). Because of the proximity of the $3s^{-1}np$ series of window resonances, SB 16 and 18 are excluded from our analysis. Fig.~\ref{Fig:Ar} (a) and (b) shows that the asymmetry parameter $\beta_2$ retrieved from the RABBIT measurements or simulations strongly oscillates as a function of delay. Experimentally, the attosecond pulse trains are intrinsically chirped. This chirp, corresponding to a  group delay dispersion of 0.018 fs$^2$ (equivalent to 130~mrad/eV), simply adds a different phase offset for different SBs and hence does not affect the PADs. In order to make the comparison with the theory easier, the chirp is removed in Fig.~\ref{Fig:Ar}(a). Theory and experiment show good qualitative agreement, with a $\beta_2$ delay dependence decreasing with energy. The difference at the minima of the oscillations might originate from a low signal to noise ratio close to 90$^\circ$, non uniform macroscopic phase effects in the gas target and limited angular resolution. 

The angle-resolved atomic delays extracted from the analysis of the experimental RABBIT traces are shown in Fig.~\ref{Fig:Ar}(c). For each data point, the SB signal was integrated over 10$^\circ$ and the oscillations were fitted according to Eq.~(\ref{Eq:RABBIT_equation}). These results are compared to our simulations and show excellent agreement, demonstrating the accuracy of both experimental and theoretical methods employed. In both cases, for each SB, the phase measured at the lowest angle was set to zero to make the comparison easier. 

In conclusion, we have generalized Fano's propensity rule to laser-assisted photoionization. The asymmetry between absorption and emission has strong implications on angle-resolved RABBIT measurements since it leads to \textit{incomplete} quantum interference. This provides a general explanation to both the delay dependence of the PADs \cite{CirelliNatCommun2018} and the angular dependence of the photoionization time delays \cite{HeuserPRA2016}. These conclusions are valid even in the case of multiple incoherent angular channels as shown by the excellent agreement between our calculations and our experimental measurements in Ar. The universality of the propensity rule implies that these conclusions can be extended to more complex atomic or molecular systems. The understanding of angular-resolved laser-assisted photoionization is essential to the study of angle-resolved attosecond dynamics in a variety of systems.

\section*{Supplemental material}

\begin{figure}[h!]
    \centering
    \includegraphics{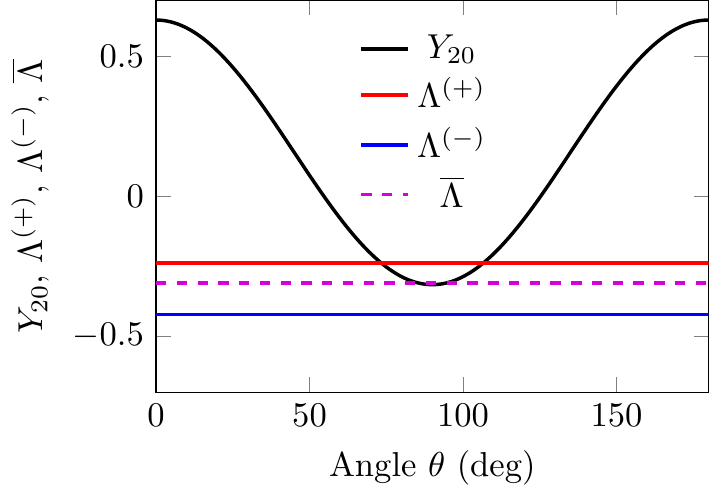}
    \caption{Spherical harmonic $Y_{20}$ and $\Lambda^{(\pm)}$, $\overline{\Lambda}$ values for which the $s$ and $d$ waves interfere destructively. These values are calculated for
    photoelectrons corresponding to SB 20 in He.}
    \label{Fig:lambdas}
\end{figure}

In the main text, we argue that the angle-dependence of the photoionization time delays can be understood with Fano's propensity rule. Here, we explain in more details how the asymmetry between absorption and emission leads to an angle-dependence of the atomic delays. For helium, neglecting the $L$-dependence of $\phi_{cc}^{(\pm)}$, the angle-resolved absorption/emission transition amplitude is given by (Eq.~8):
\begin{equation}
\!\!\!\!A^{(\pm)}(\theta)\!\approx\!e^{i[\pm\omega\tau+\eta_\lambda^{(\pm)}+\phi_{cc}^{(\pm)}]}\left[\left| M_{120}^{(\pm)}\right|Y_{20}(\theta)\!-\!\left|M_{100}^{(\pm)}\right|Y_{00}\right].
    \label{Eq:AThetaHe}
\end{equation}
The right factor changes sign when  $Y_{20}(\theta)$ crosses  $\Lambda^{(\pm)}=Y_{00}|M_{100}^{(\pm)}|/|M_{120}^{(\pm)}|$ and the argument of $A^{(\pm)}(\theta)$ undergoes a $\pi$ phase jump. At these angles, the function $\left|A^{(\pm)}(\theta)\right|^2$ goes to zero, which can be interpreted as destructive interference between the two final states. Fig.~\ref{Fig:lambdas} shows that $Y_{20}(\theta)$ crosses $\Lambda^{(+)}$ at two angles $75^\circ$ and $105^\circ$, while $\Lambda^{(-)}$ never crosses $Y_{20}$. As a consequence, the angle-dependent time delay presents a sharp variation close to the position of the destructive interference in the absorption path as seen in Fig.~3 in the main text. As the energy of the photoelectron increases, the asymmetry between absorption and emission is reduced and $\Lambda^{(+)}$ and $\Lambda^{(-)}$ get closer.

If the absorption and emission paths were equivalent, $\Lambda^{(+)}$ and $\Lambda^{(-)}$ would coincide. We can assume that  $\Lambda^{(\pm)}=\overline{\Lambda}=[\Lambda^{(+)}+\Lambda^{(-)}]/2$. In this case, both absorption and emission paths undergo $\pi$ phase jumps at the same angles, $86^\circ$ and $94^\circ$, {\it therefore cancelling each other} and leading to an angle-independent sideband phase. 
The angular dependence of photoionization time delays can hence be attributed to the asymmetry between absorption and emission, which comes from the generalized Fano's propensity rule. 

\begin{acknowledgments} 
The authors acknowledge Linnea Rading for designing the VMIS. The authors acknowledge support from the Swedish Research Council (Grants No. 2013-08185, 2014-3724, 2016-03789), the European Research Council (advanced grant PALP-339253) and the Knut and Alice Wallenberg Foundation.  
\end{acknowledgments}


%

\end{document}